\documentclass[10pt,preprint]{emulateapj}
\slugcomment{Accepted by ApJL}
\usepackage{amsmath}
\usepackage{placeins}
\usepackage{hyperref}
\usepackage{natbib}

\newcommand{\Msol}{\hbox{$M_{\odot}$}}

\newcommand{\mdotyr}{\hbox{$M_\odot$ yr$^{-1}$}}
\newcommand{\mdot}{\hbox{$\dot{M}$}}

\newcommand{\HST}{\textit{HST}}

\shorttitle{Accretion onto very low mass companions of young stars}
\shortauthors{Zhou et al.}

\begin{document}

\title{Accretion onto planetary mass companions of low-mass young stars}

\author{Yifan Zhou$^{1,2}$, Gregory J. Herczeg$^{1,2}$, Adam
  L. Kraus$^{3}$, Stanimir  Metchev$^{4,5}$ and Kelle L. Cruz$^{6,7}$}

\altaffiltext{1}{Kavli Institute for Astronomy and Astrophysics,
  Peking University, Yi He Yuan Lu 5, Haidian Qu, Beijing 100871,
  People's Republic of China}
\altaffiltext{2}{Department of Astronomy, School of Physics, Peking
  University, Yi He Yuan Lu 5, Haidian District, Beijing 100871,
  P.R. China}

\altaffiltext{3}{Department of Astronomy, The University of Texas at
  Austin, Austin, TX 78712, USA}

\altaffiltext{4}{Department of Physics \& Astronomy, The University of Western Ontario, London, ON N6A 3K7, Canada}
\altaffiltext{5}{Department of Physics and Astronomy, State University of New York,
Stony Brook, NY 11794, USA }
\altaffiltext{6}{Hunter College, City University of New York, Dept. of
Physics and Astronomy, New York, NY 10065}
\altaffiltext{7}{American Museum of Natural History, Astrophysics Dept, New York, NY 10025}
\begin{abstract}
  Measurements of accretion rates onto planetary mass objects may
  distinguish between different planet formation mechanisms, which
  predict different accretion histories.  In this Letter, we use
  \HST/WFC3 UVIS optical photometry to measure accretion rates onto
  three accreting objects, GSC06214-00210 b, GQ Lup b, and DH Tau b,
  that are at the planet/brown dwarf boundary and are companions to
  solar mass stars.  The excess optical emission in the excess
  accretion continuum yields mass accretion rates of $10^{-9}$ to
  $10^{-11}$ \Msol/yr for these three objects.  Their
  accretion rates are an order of magnitude higher than expected from
  the correlation between mass and accretion rates measured from the
  UV excess, which is applicable if these wide planetary mass companions
  formed by protostellar core fragmentation.
The high accretion rates and large seperation from
  the central star demonstrate the presence of massive disks around these
  objects. Models for the formation and evolution of wide planetary mass
  companions should account for their large accretion rates. High
  ratios of H$\alpha$ luminosity over accretion luminosity for objects
  with low accretion rates suggest that searches for H$\alpha$
  emission may be an efficient way to find accreting planets.
\end{abstract}
 

%
%


\section{INTRODUCTION}

In recent years, many young objects with planetary masses have been
directly imaged at distances of tens to a few hundred AU from the central
star \citep[e.g.][]{Neuhauser2005,Itoh2005,Kraus2008,Marois2008,Kraus2012,Quanz2013}.
The formation of these planetary mass companions may occur
by planetary core formation followed by vigorous gas accretion, by 
gravitational instabilities within the disk, or by uneven fragmentation of
the collapsing protostellar core. These formation mechanisms each predict
different accretion histories, which lead to different contraction
timescales and luminosities at ages $<100$
Myr \citep{Spiegel2012,Mordasini2013}.  
Since younger objects are brighter and therefore more easily detected,
the uncertain formation and accretion history leads
to significant uncertainties in derived masses of directly imaged exoplanets.

If very low mass companions have similar disk fractions as solar
mass stars and have accretion rates that are expected for their mass,
then their formation may be consistent with the low-mass tail of the IMF
resulting from protostellar core fragmentation.  A similar approach established
that free floating brown dwarfs likely form by the collapse of a protostellar
core, similar to solar mass stars
\citep[e.g.][]{White2003,Muzerolle2003,Mohanty2005,Joergens2013}.  On
the other hand, planetary mass objects that form by core accretion or
by disk gravitational instabilities may have different accretion and
disk properties.  Core accretion models for giant planet formation
predict that for a few Myr, the accretion rates of planetary
companions should be very large and may even dominate the flux for a
few Myr \citep{Spiegel2012,Mordasini2013}.

Previously, accretion has been detected onto two very low mass
companions of solar mass stars based on emission in the
Paschen-$\beta$ line
\citep[e.g.][]{Seifahrt2007,Bowler2011,Bonnefoy2013}.  The line
luminosities may then be converted to accretion rate using established
correlations \citep{Natta2006}, however the uncertainties and scatter 
in the correlations
introduce significant uncertainties in the resulting accretion rates,
especially beyond the mass accretion rate regime over which the
correlation was calculated. In this {\it Letter}, we use broadband
optical HST/WFC3 photometry to directly measure the accretion rate of
very low-mass companions to the pre-main sequence stars GSC
06214-00210, GQ Lup, and DH Tau.  The accretion luminosity is measured
directly from the excess line and continuum emission, following the
approach of \citet{Herczeg2008}.  The use of photometry to obtain
accretion rates is similar to the approach of \citet{Hartmann1998} and
\citet{White2001}.

\begin{table}[!ht]
\centering
\caption{Observations}
\label{tab:1}
\begin{tabular}{lcccc}
  WFC3 & $t_{exp}$& Mag & Flux$^a$ &Error\\
  Filter & s &~& (erg cm$^{-2}$ s$^{-1}$) & \\
  \hline
  \multicolumn{5}{c}{GSC6214-210 b}\\
  \hline
  F336W   &2400&21.9&    5.55E-18       &   1.4\%   \\  
  F390W   &480&22.1&    8.16E-18        &  1.6\%   \\      
  F475W   &320&21.9&    4.23E-18        & 5.5\%   \\      
  F555W   &240&22.3&    4.62E-18        &  6.2\%   \\      
  F625W   &140&20.8&    1.13E-17        &  3.2\%   \\      
 (F625W)$^{b}$  &~&  21.3&    7.33E-18        &  3.2\% \\
  F656N   &2380&15.7&       7.08E-16   &       0.35\%   \\      
  F673N   &800&21.5&    4.66E-18        &  7.2\%   \\      
  F775W   &80&20.2&    1.05E-17        & 2.9\%   \\       
  F850LP  &50&18.4&     3.32E-17       &   2.0\%   \\      
  \hline
  & \multicolumn{2}{l}{$\log L_{slab}/L_\odot$} & \multicolumn{2}{c}{-4.65}\\
  & \multicolumn{2}{l}{$\log L_{H-\alpha}/L_\odot$} & \multicolumn{2}{c}{-5.03}\\
  & \multicolumn{2}{l}{EW$_{H\alpha}$} & \multicolumn{2}{c}{$1600$ \AA}\\
  \hline
  \hline
  \multicolumn{5}{c}{GQ Lup b}\\
  \hline
  F336W   &400& 19.2 &    6.77E-17  &    1.6\%   \\      
  F390W   &50&  20.1 &    5.42E-17  &   2.0\%   \\      
  F475W   &40&  20.0 &    2.61E-17  &   5.2\%   \\      
  F555W   &30&  20.2 &    3.19E-17  &   3.5\%  \\      
  F625W   &20&  18.9 &    6.88E-17  &   8.0\%   \\      
 (F625W)  &~&   19.2 &    5.03E-17  &   8.9\%   \\
  F656N   &250& 15.9 &    5.92E-16  &    3.9\%   \\      
  F673N   &160& 19.0 &    4.84E-17  &    15.3\%  \\      
  F775W   &40&  17.8 &    9.60E-17  &   2.9\%   \\    
  F850LP  &20&  16.2 &    2.75E-16  &    1.8\%   \\      
  \hline
  & \multicolumn{2}{l}{$\log L_{slab}/L_\odot$} & \multicolumn{2}{c}{-2.91}\\
  & \multicolumn{2}{l}{$\log L_{H-\alpha}/L_\odot$} & \multicolumn{2}{c}{-4.69}\\
  & \multicolumn{2}{l}{EW$_{H\alpha}$} &\multicolumn{2}{c}{$180$ \AA}\\
  \hline
  \hline
  \multicolumn{5}{c}{DH Tau b}\\
  \hline
  F336W  &1400&24.2&6.90E-19&30\%\\
  F390W  &360&24.9&6.62E-19&20\%\\
  F475W  &280&24.3&4.85E-19&9.0\%\\
  F555W  &160&24.5&5.98E-19&7.2\%\\
  F625W  &100&23.0&1.51E-18&5.7\%\\
 (F625W) &~&23.1&1.35E-18&6.6\%\\
  F656N  &1200&19.0&3.48E-17&3.4\%\\
  F673N  &500&23.0&1.23E-18&21\%\\
  F775W  &80&20.2&1.07E-17&2.7\%\\
  F850LP&40&18.0&5.16E-17&1.8\%\\
  \hline
  & \multicolumn{2}{l}{$\log L_{slab}/L_\odot$} & \multicolumn{2}{c}{-5.40}\\
  & \multicolumn{2}{l}{$\log L_{H-\alpha}/L_\odot$} & \multicolumn{2}{c}{-6.19}\\
  & \multicolumn{2}{l}{EW$_{H\alpha}$} &\multicolumn{2}{c}{$450$ \AA}\\
  \hline
  \multicolumn{5}{l}{$^a$Observed fluxes not corrected for extinction.}\\
  \multicolumn{5}{l}{~~Listed luminosities are corrected for
    extinction.}\\
  \multicolumn{5}{l}{$^{b}$ F625W fluxes in () are fluxes with
    H$\alpha$}\\
  \multicolumn{5}{l}{~~emission subtracted.}
\end{tabular}
\end{table}

\begin{table*}[t]
\centering
\caption{Sample Properties}
\label{tab:2}
\begin{tabular}{lcccccccccccc}
  \hline
  Object & Sep. ($^{\prime\prime}$) & PA (deg) & d (pc) & $A_V$ &
  $T_{{\rm eff}}$ (K)& $\log L_{phot}$ & $R$ ($R_{J}$) &
  $M(M_{J})$&$\log L_{acc}$ & $\log \dot{M}$ & Ref.\\
  \hline
  GSC 06214&  $2.19\pm0.01$ & $175.2\pm0.2$ & $145\pm 15$ & 0.2  & 2700$\pm
  200$ & $-3.1\pm0.1$& $1.5\pm0.5$ &14$\pm 2$&  -4.7 & -11.0
  & 1\\
  -00210 b&&&&&2200 &-3.16&$1.8\pm0.5$&15$\pm 3$& -4.6 & -10.8  & - \\
  \hline
  GQ Lup b & $0.713\pm0.006$ & $-83.6\pm0.7$ & $155\pm 15$ & 1.5 &
  $2050\pm350$ &-2.25$\pm0.24$&6.5$\pm 2$&24$\pm12$& -2.9 & -9.3 & 2,3\\
  &&&&&2400 &-2.24&$4.6\pm1.4$&31$\pm10$& -2.9 & -9.3 & - \\
  \hline
  DH Tau b &  $2.31\pm0.02$ & $138.5\pm0.1$ & $145\pm15$ & 0.7 &
  $2350\pm150$&-2.70$\pm0.11$& $2.6\pm0.7$ &11$^{+10}_{{-3}}$& -5.4 & -11.5 & 2,4\\
  &&&&&2200 &-2.82&$2.7\pm0.8$&11$\pm3$& -5.3 & -11.3 & - \\
  \hline
  \multicolumn{12}{l}{For each object, the first line refers to literature
    $T_{eff}$, $L_{phot}$, $R$, and $M$.  The second line is measured
    here.}\\
  \multicolumn{12}{l}{~~~The $L_{acc}=L_{slab}+L_{{\rm H}\alpha}$ and $\dot{M}$ are measured for
    the photospheric parameter corresponding to each line.}\\
  \multicolumn{12}{l}{For a description of the uncertainty in $\dot{M}$ see \S 3}\\
  \multicolumn{12}{l}{References: (1)\cite{Bowler2011}
    (2)\cite{Patience2012}, (3)\cite{Seifahrt2007}, (4)\cite{Itoh2005}}\\
\end{tabular}
\end{table*}

\section{OBSERVATIONS}
We used {\it HST}/WFC3 UVIS2 to obtain optical photometry of GQ Lup b
on 25 Feb.~2012, GSC 06214-00210 b on 22 Feb~2012, and DH Tau b on
22~Jan 2012 in HST program GO 12507 (P.I. A.~Kraus).  The high spatial
resolution of HST is typically needed to resolve the compnents at
optical wavelengths.  The reduced and flatfielded images were
downloaded from the MAST archive for analysis.  The observation log
and extracted fluxes are listed in Table~\ref{tab:1}.

\subsection{Sample Selection and Properties}
The total dataset consists of optical imaging of the 12 planetary mass
companions to young solar mass stars that had been identified as of
Feb.~2011, when the proposal was submitted.  In this paper, we analyze
three objects, GQ Lup b \citep{Neuhauser2005}, GSC 06214-00210 b \citep{Kraus2008}, and DH
Tau b \citep{Itoh2005}, that show excess emission in H$\alpha$ and in
the U-band.  A subsequent paper will describe the full sample.
Emission in Pa$\beta$, an accretion diagnostic, has previously been
detected from GSC 06214-00210 b and DH Tau b
\citep{Bowler2011,Bonnefoy2013}.  Pa$\beta$ emission was detected in
one of three near-IR spectra of GQ Lup b
\citep{Seifahrt2007,McElwain2007,Lavigne2009}.

The properties of the three planetary mass companions are listed in
Table~\ref{tab:2}.  The extinctions of the companion are assumed to be
equal to that of the primary. The extinctions to the accreting stars
GQ Lup and DH Tau are calculated from optical spectra by
\citet{Herczeg2014}. The extinction to the non-accreting star
GSC06214-210 is calculated here from the R-J color. Extinction curves
applied here assume $R_V=3.1$ \citep{Fitzpatrick1990}.  The
distances and ages of GSC06214-00210, GQ Lup and DH Tau are assumed to
be that of their parent clouds: Upper Sco OB Association at 145 pc
\citep{Zeeuw1999} and 5--12 Myr \citep{Preibisch2002,Pecaut2012},
Lupus 1 at 155 pc \citep{Lombardi2008} and 3 Myr \citep{Alcala2013},
and Taurus at 145 pc \citep{Torres2009} and 1--2 Myr
\citep[e.g.][]{Luhman2004}, respectively.  The listed separations and
position angles are calculated from the median of our measurements for
each filter with the plate scale of 0.040 arcsec/pixel.

\subsection{Source Extraction}
\begin{figure}[!tb]
\centering%
\plottwo{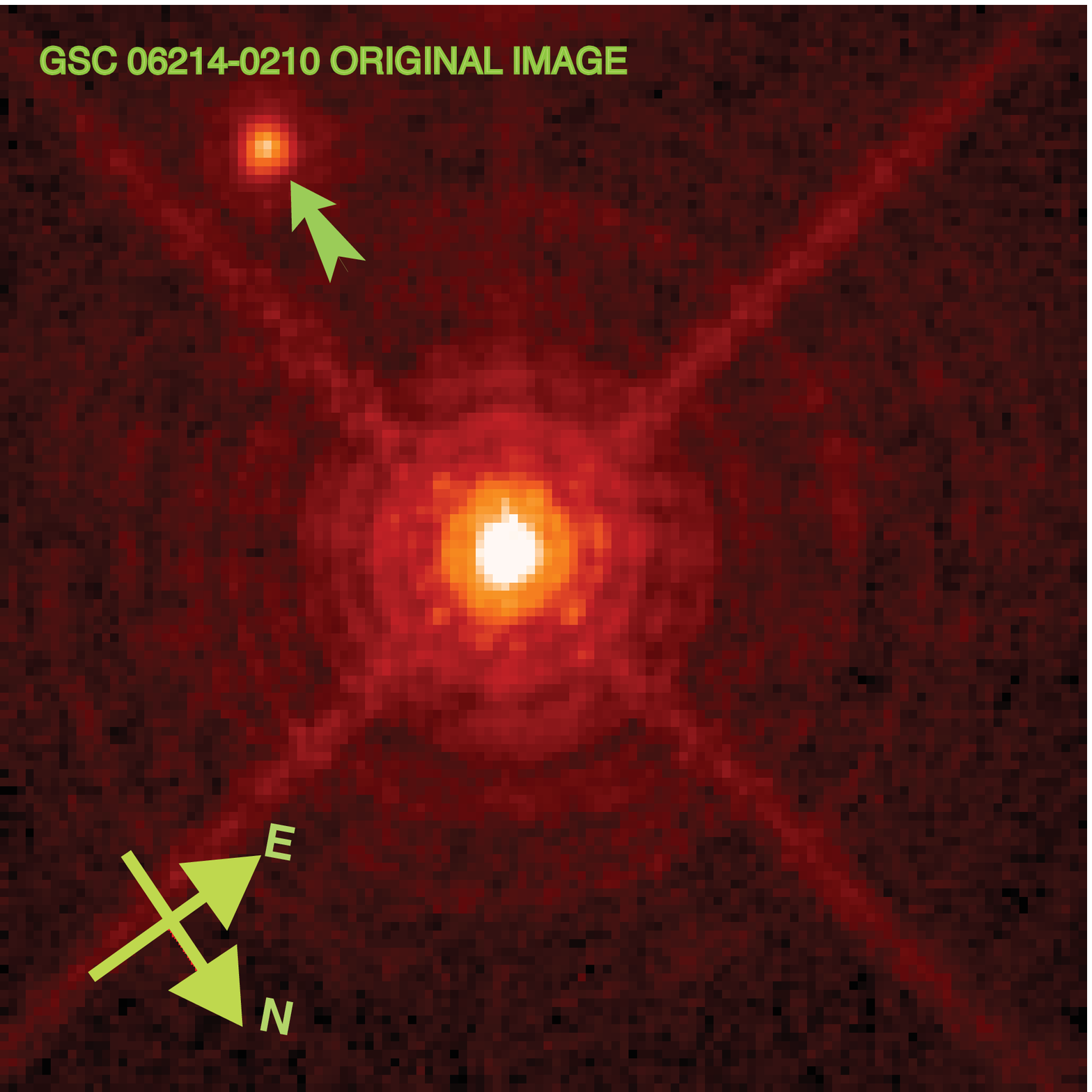}{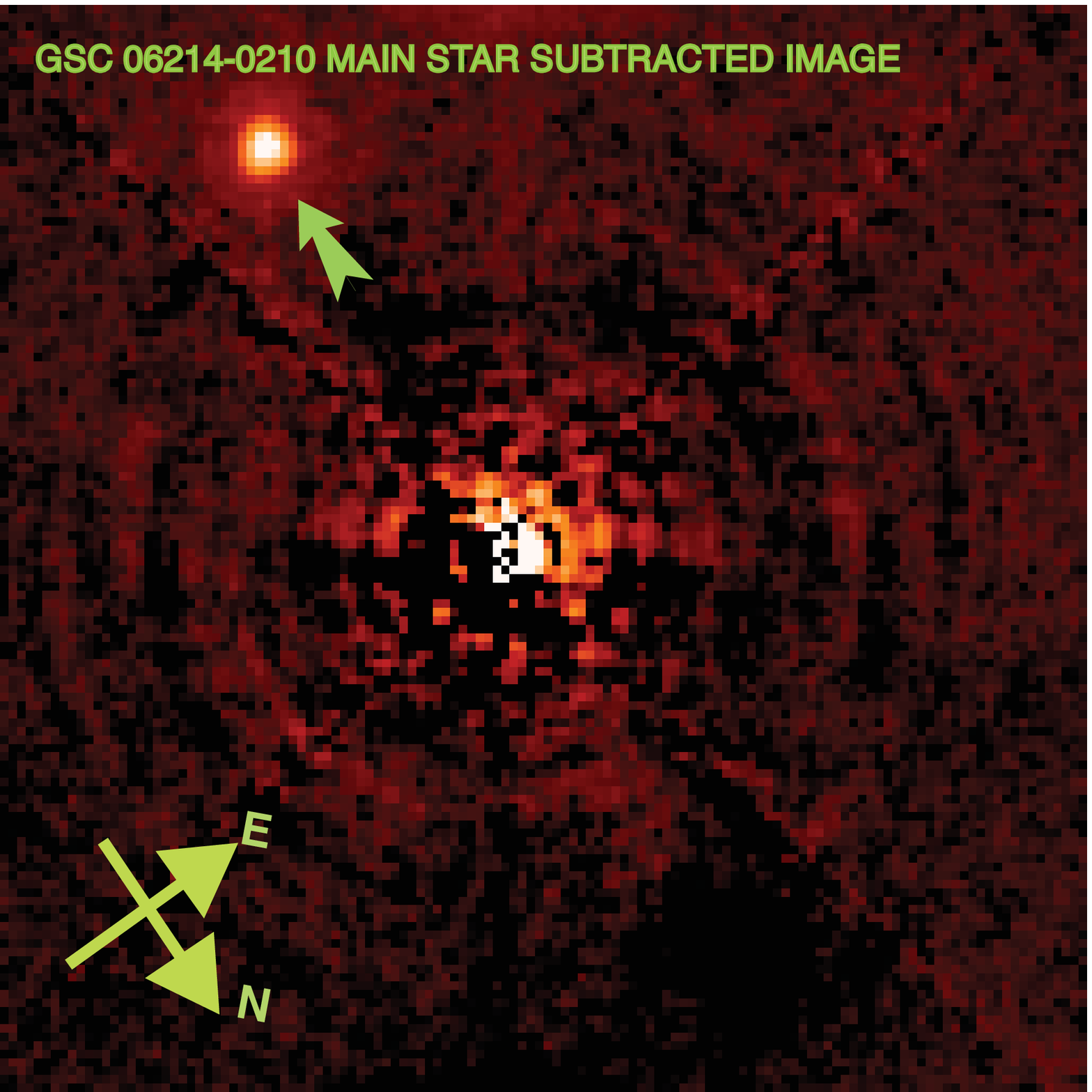}\\
\plottwo{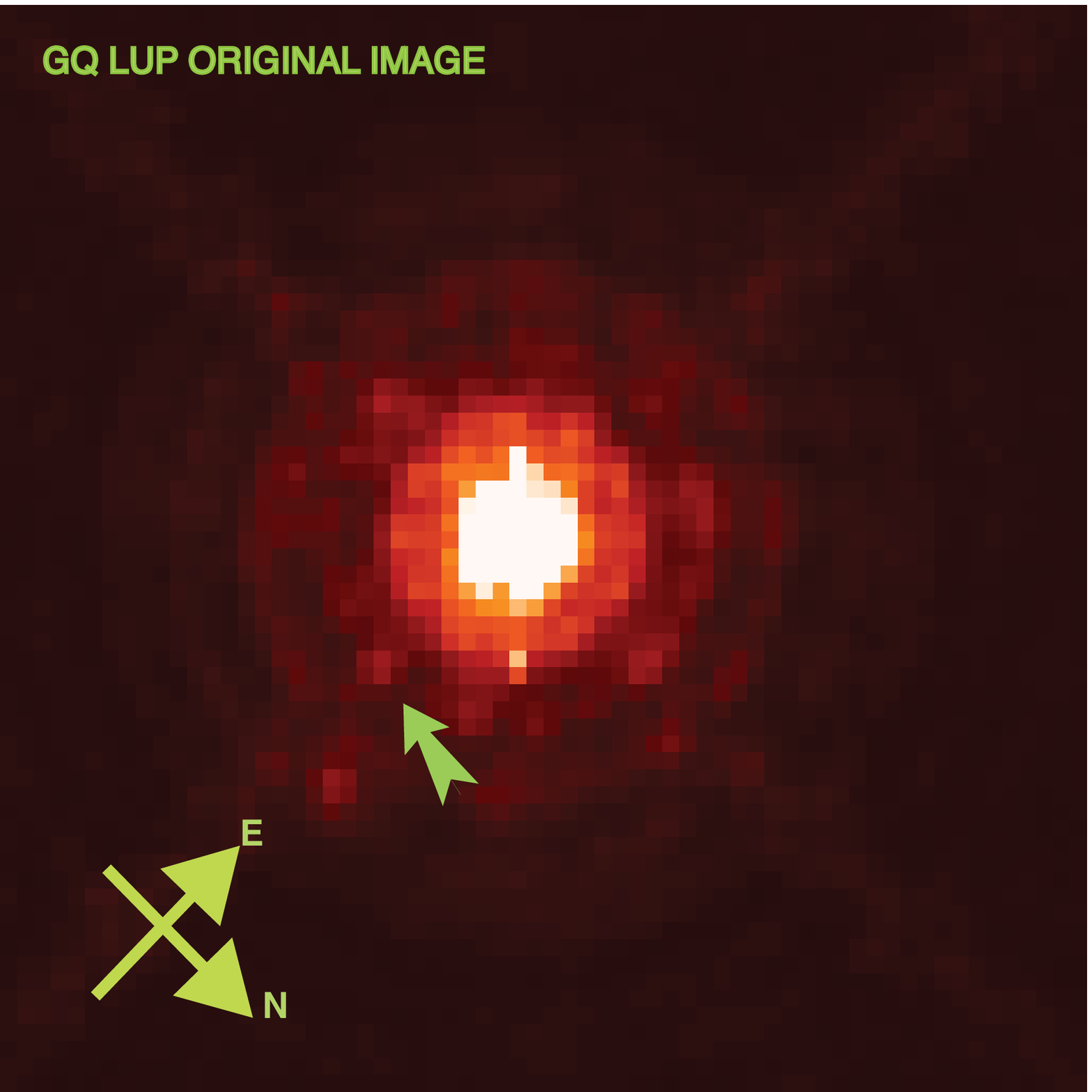}{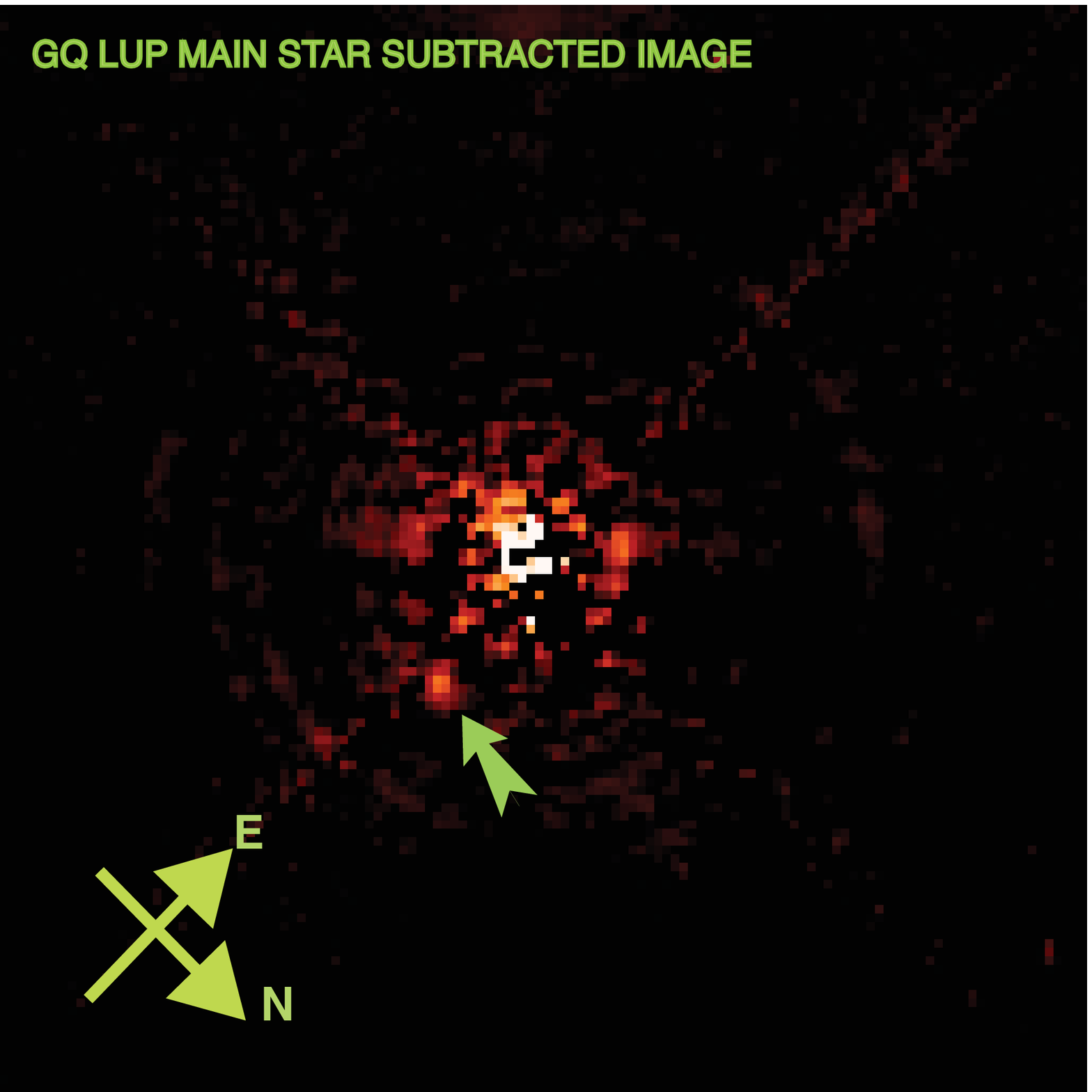}\\
\plottwo{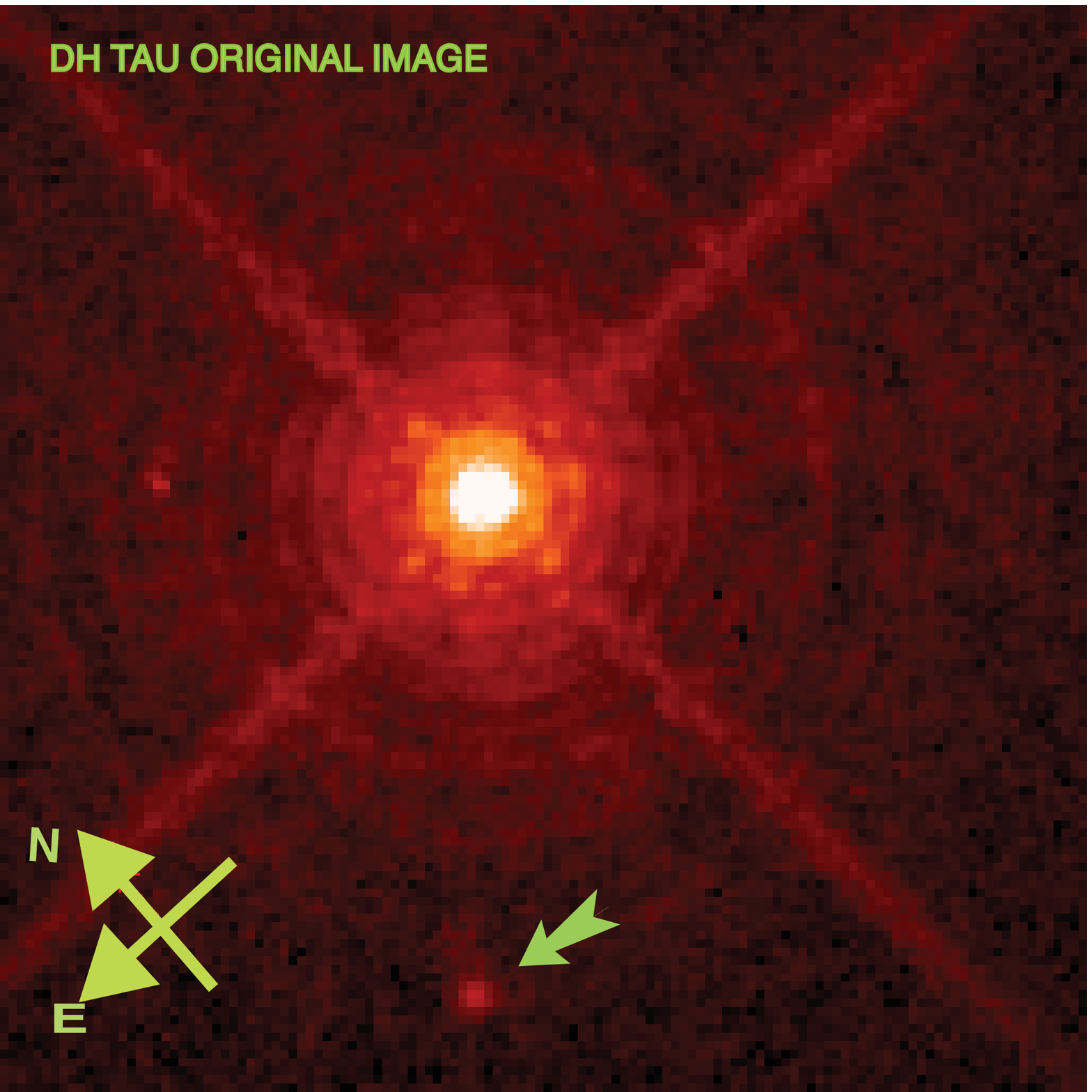}{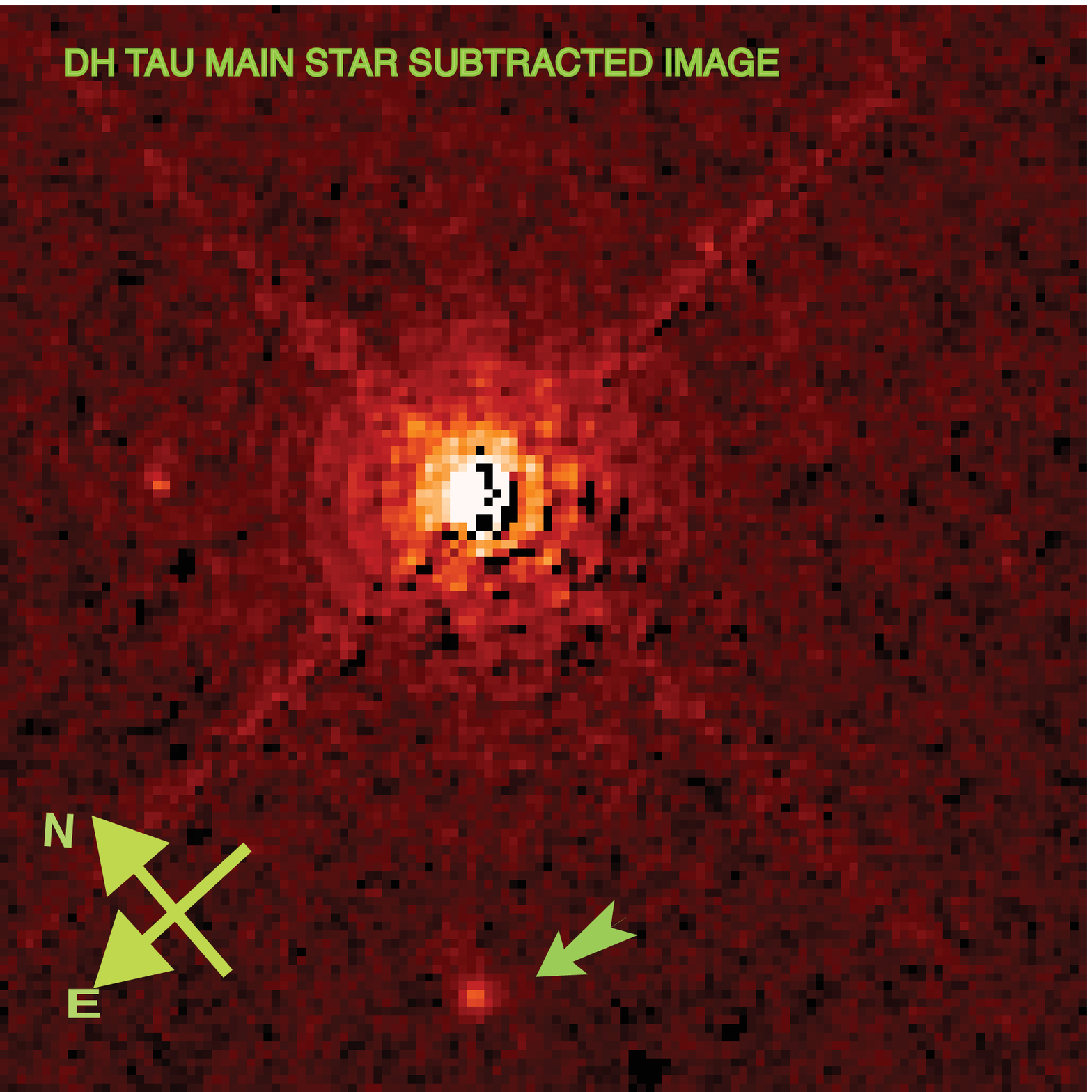}\\

\caption{Subtracting the primary in the GSC 06214-00210 F656N image
  (top), the GQ Lup image (middle) F656N and DH Tau (bottom) F656N.
  For GSC06214-00210 and DH Tau, the primary star is subtracted by a
  scaled PSF. The secondary is then fit with a model psf For GQ Lup,
  The primary star from the original image (left) is rotated by 180
  degrees and subtracting, leaving a clearer image of the
  secondary. The bright point on the left of the primary star of DH
  Tau only appears on the image of F656N.}
\label{fig:1}
\end{figure}

The images consist of bright emission from the primary star and faint
emission from the companion. Isolated stars in the field are used
to synthesize a template point spread function (psf) in each
image. For the GSC 06214-00210 images obtained with the F656N, F673N,
F850LP filters, the template psf is first scaled to the primary and
subtracted from the image, leaving only the faint companion. For other
images of GSC 06214-00210 A and all the images of GQ Lup A and DH Tau
A, the primary stars are heavily saturated and poorly fit with the
template.  The saturated spot is assumed to be symmetric about
the x and y axis and rotate the stellar image to subtract out the
primary.  Fig.~\ref{fig:1} demonstrates our psf subtraction for
the F656N observations of all three objects.

After the primary flux is subtracted, the flux in the secondary
component measured by fitting the spot with the template psf.  The
template brightness was measured from aperture photometry and
converted to a flux based on the aperture correction and flux
conversion listed in the WFC3
manual\footnote{http://www.stsci.edu/hst/wfc3/documents/handbooks/\\currentIHB/c06\_uvis07.html\#391868}.
The extraction apertures have radii as large as 30 pixels ($1\farcs2$)
pixels for images with bright secondaries well separated from the
primary, to as little as 3 pixels ($0\farcs12$) for faint images.  The
fluxes of the psfs are then corrected by applying the flux-aperture radius curve
from the WFC3 manual.

The photometry of the companion is then calculated by fitting the psf
template to the image of the companion.  GQ Lup b and GSC06214-0210 b
are located close to the diffraction spikes, which are avoided in the
2D fits to the image.  The uncertainties are calculated from
statistical error and fluctuation of the background in an annular
region around the primary star.

The final photometric results and $1 \sigma$ uncertainties are listed
in Table~\ref{tab:1}.  The H$\alpha$ fluxes are calculated by
convolving a flux above the model fit with the filter transmission
curve to obtain the measured flux.  The approximate equivalent widths
are calculated by dividing the measured H$\alpha$ flux by the average
flux in the F625W and F673N filters.

A faint object is detected in only the F656N image of DH Tau, with
flux of $2.4\times 10^{-17}$ ergs/cm$^{2}$/s/\AA, a
separation of $1\farcs53$ and a position angle of 46.37$^{\circ}$ to
relative to the primary star.  This object has upper limits of
$2.1\times 10^{{-19}}$ and $3.7\times10^{-19}$ ergs/cm$^{2}$/s/\AA\
in the F775W and F850LP filters. A faint object is also found at a similar
relative position to DI Tau in the same image, indicating that both
detections are likely artifacts.\par

\begin{figure*}[!ht]
  \centering
\plottwo{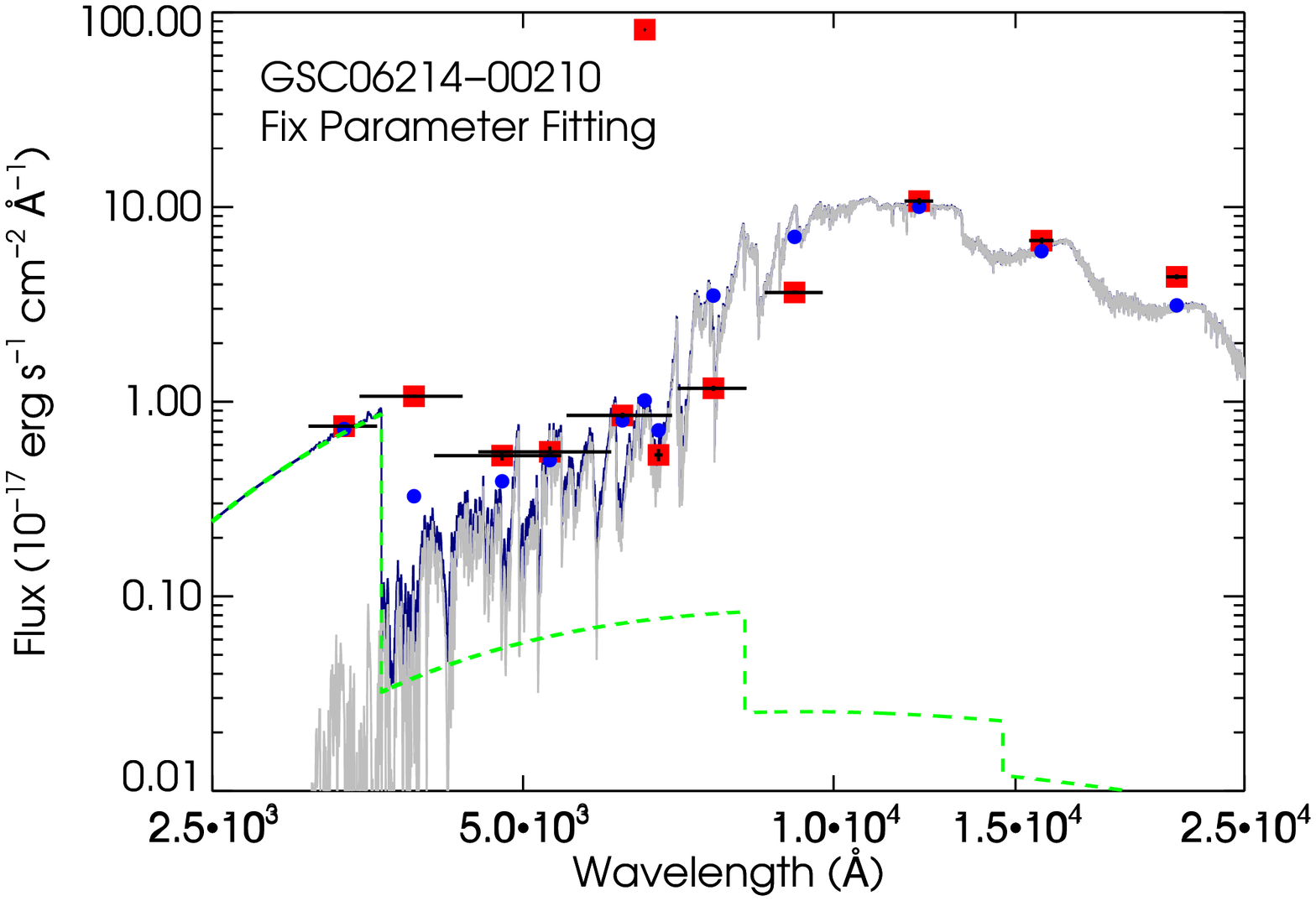}{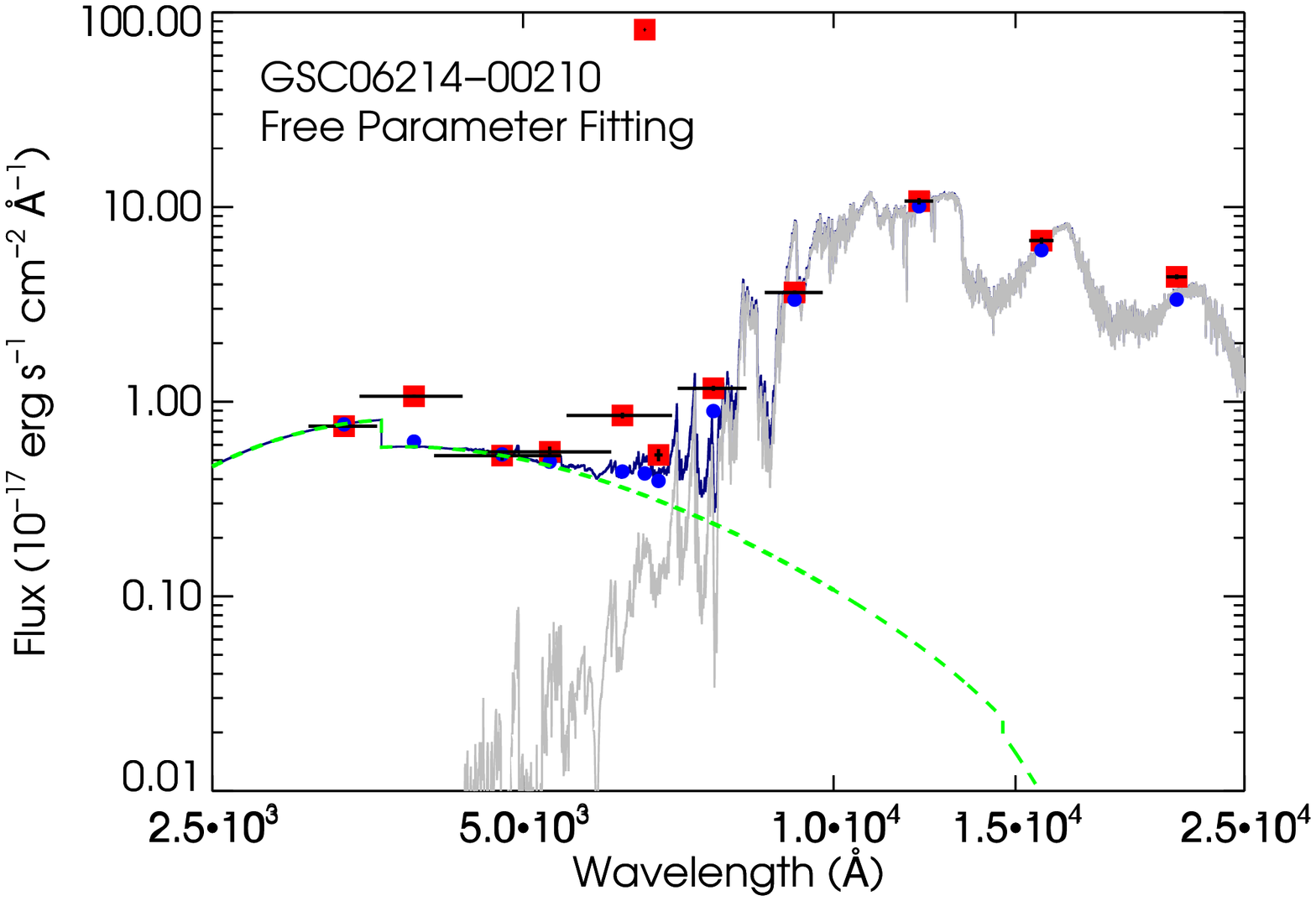}
\plottwo{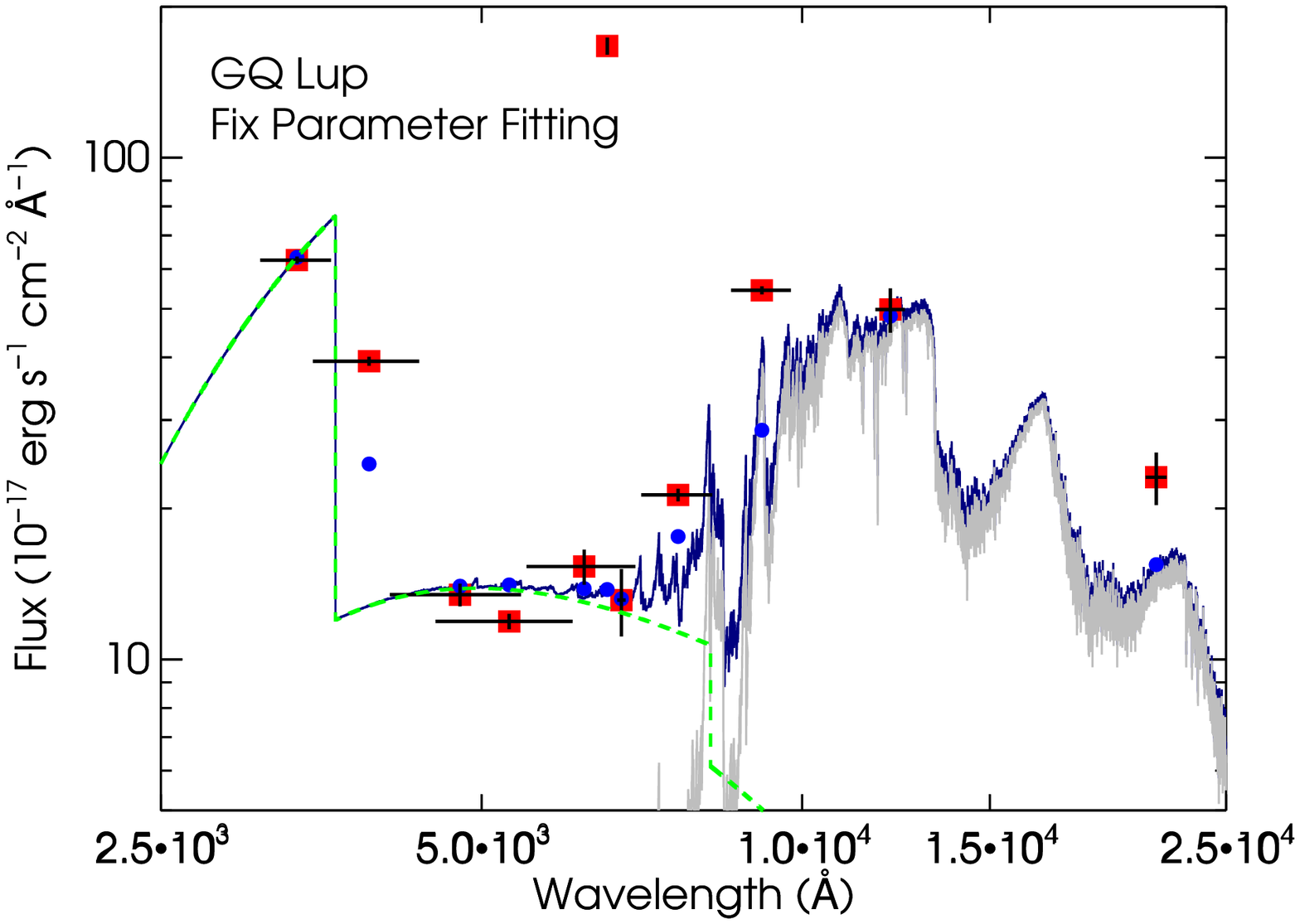}{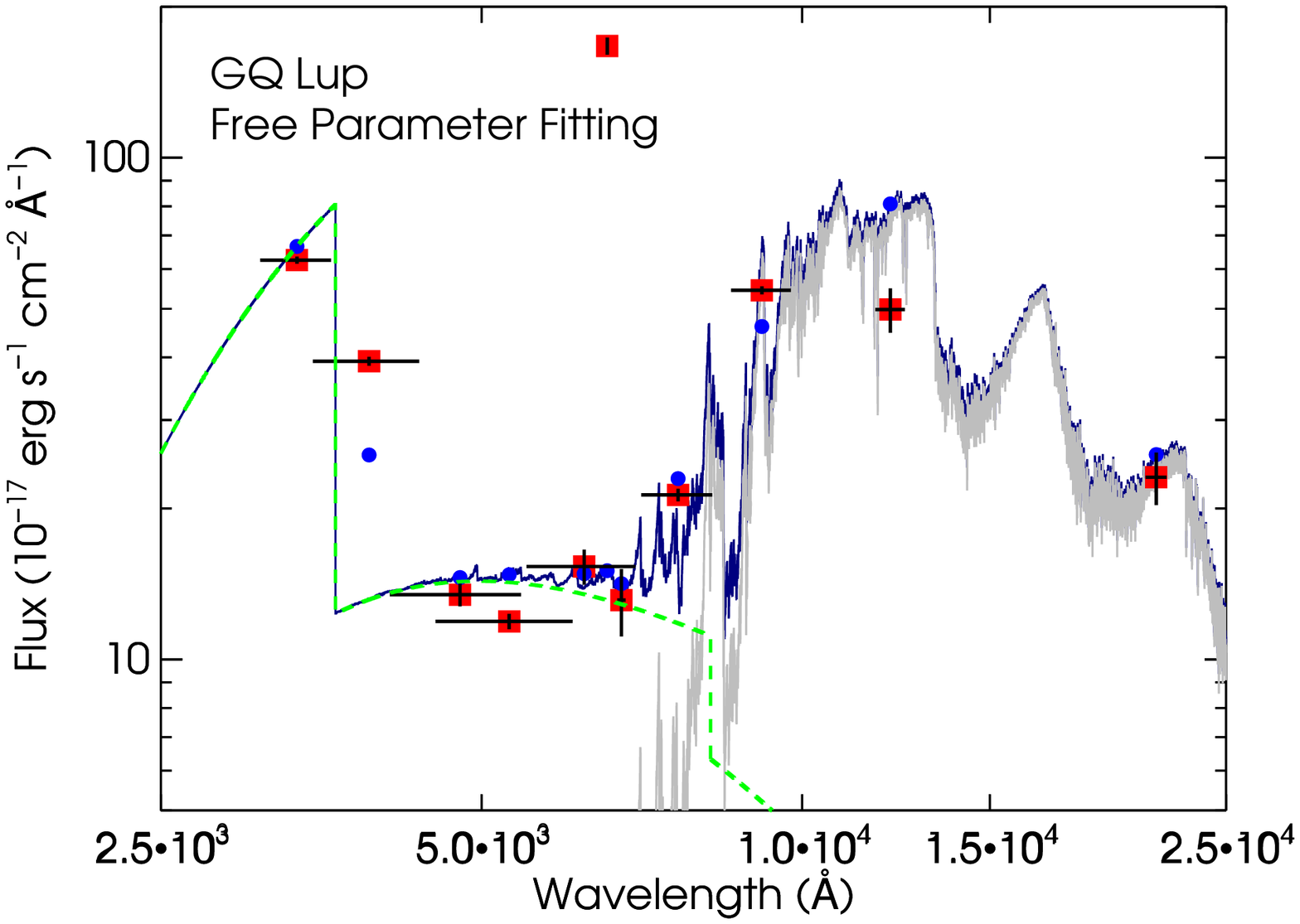}
\plottwo{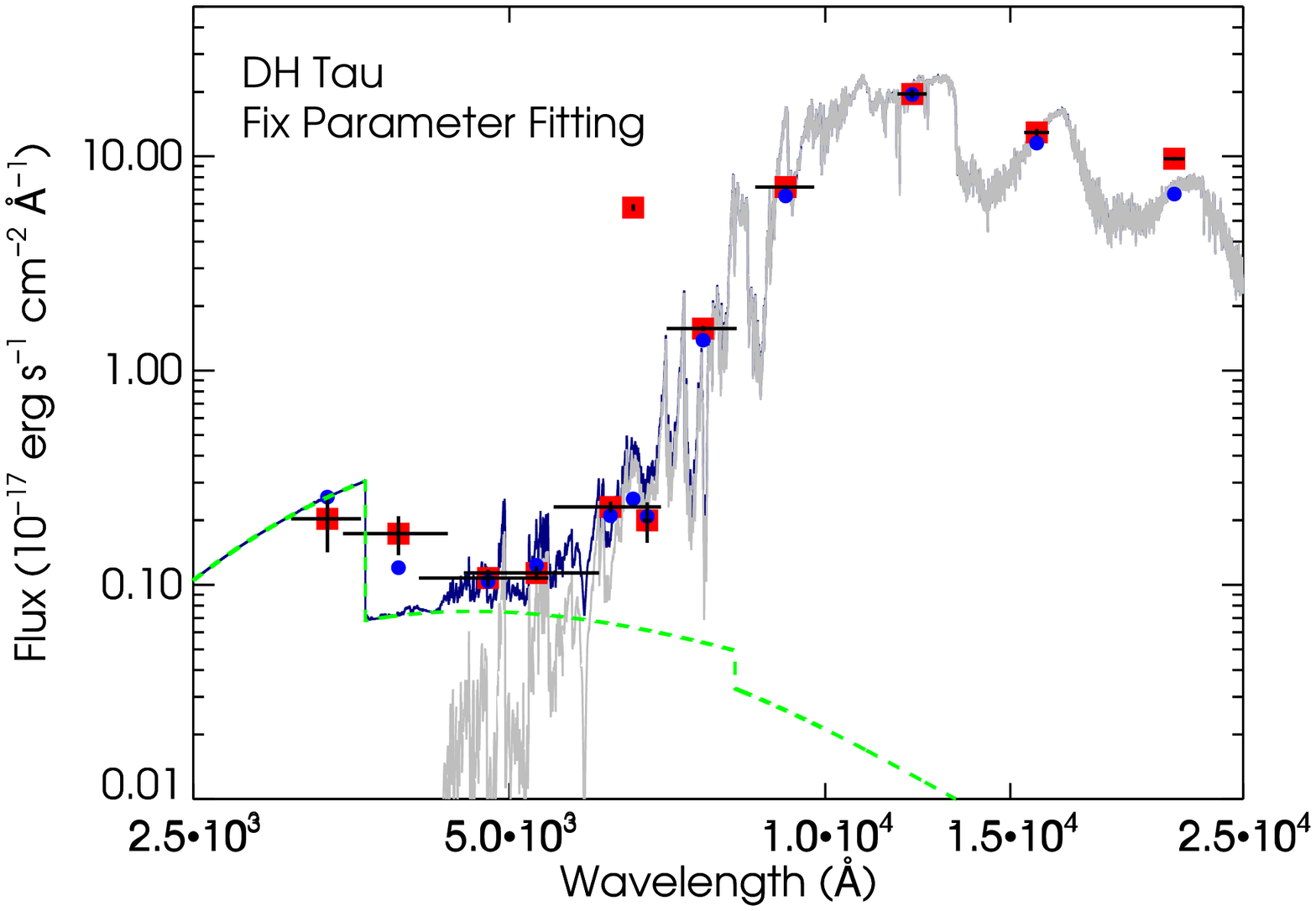}{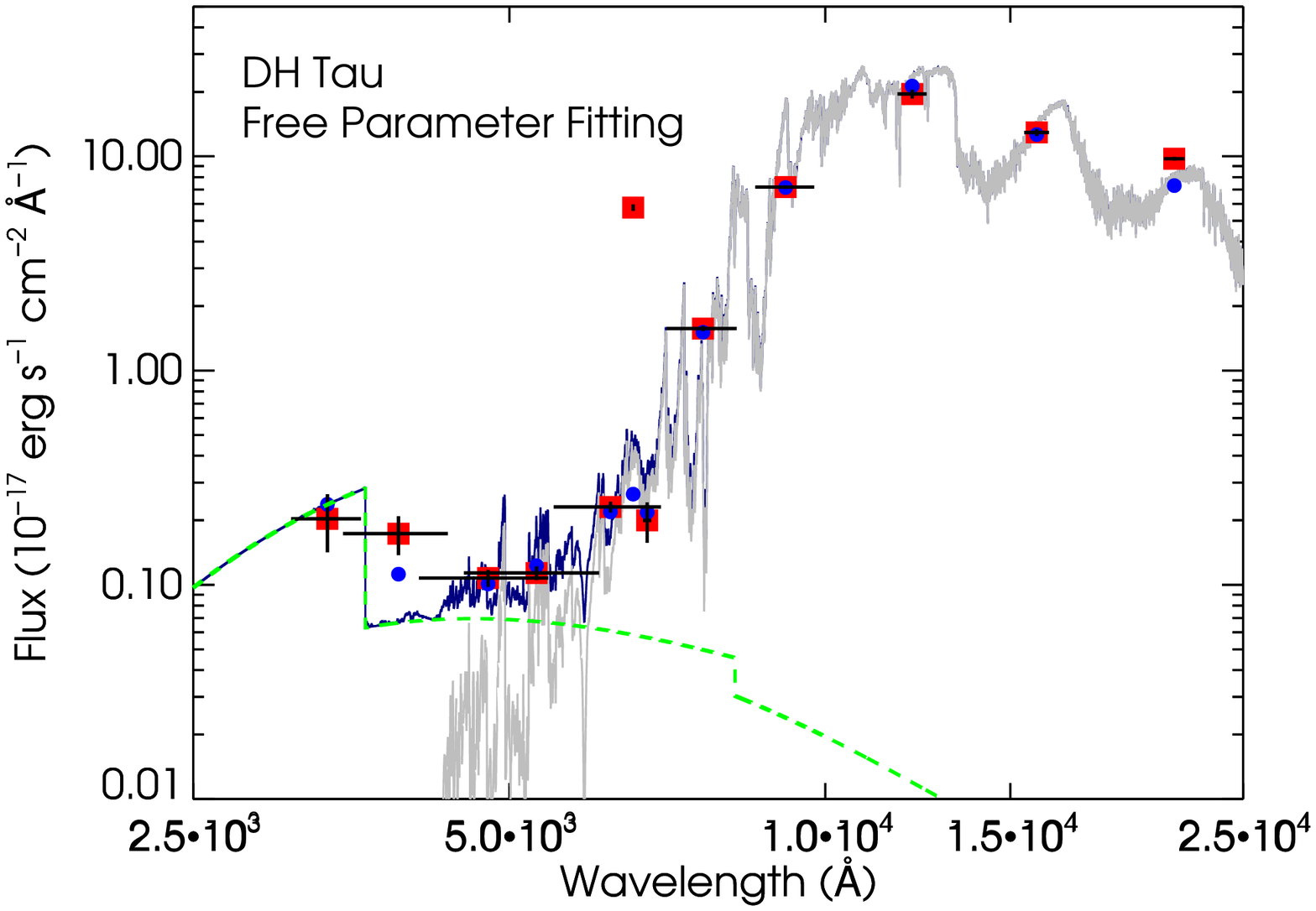}
\caption{Optical and NIR SED with BT-settl grid and slab emission of
  GSC06214-00210 (top), GQ Lup b (middle), and DH Tau b (bottom).  The
  left figures are the results for fixed parameter fitting. The right
  figures are for free parameter fitting. Red square points are
  observed photometry results. Green dashed lines are emission from a
  pure hydrogen layer, gray lines are the fluxes from BT-settl model
  spectra ,dark blue lines are the total flux of photosphere and
  accretion model. The integrate flux of the models are plotted in
  blue points. The SEDs present clear optical excess
  emissions. H$\alpha$ emissions in F625W filter have been
  subtracted.}
\label{fig:2} 
\end{figure*}

\section{Results}

Fig.~\ref{fig:2} shows our optical photometry combined with JHK
photometry of GQ Lup and DH Tau from \citet{Patience2012} and
GSC0~6214-00210 from \citet{Bowler2011}.  The sources
are all brightest at red wavelengths, as expected for very low mass
sources.  However, the optical fluxes are relatively flat with
wavelength, and higher in the F330W and F390W filters than at longer
wavelengths.  The emission in the narrow-band F656N filter is much
brighter than the nearby spectral regions because of H$\alpha$
emission, which also contributes 10\%, 25\%, and 77\% of the photons
in the broad-band F625W filter for GQ Lup, FW Tau, and GSC~06214-0210,
respectively.

The spectrophotometry is explained as the combination of a photosphere
from a very low mass object and an accretion spectrum.  Accretion is
characterized by excess optical emission in the H Paschen continuum, a
stronger excess in the H Balmer continuum shortward of 3700 \AA, and
strong emission in H lines \citep[see, e.g.,][]{Calvet1998,Herczeg2009}.  The
brightness of emission in the F330W filter is interpreted as the
excess Balmer continuum emission.  Emission in high Balmer lines and
\ion{Ca}{2} H \& K likely explains the bright emission in the F390W
filter, relative to longer wavelengths.  
Other optical lines may be strong enough
to contaminate emission in other filters \citep[e.g.][]{Bowler2014}, but
are assumed to be negibible.  In particular, the F673N filter is
consistent with expectations of the photospheric flux, indicating at
most a minimal contribution from the [S II] 6725 \AA\ doublet.

The photometry is modeled as the sum of the object photosphere
and an accretion spectrum. Synthetic photometry for the photosphere
is obtained from the BT-Settl grid with CIFIST opacities
\citep{Allard2012}.  The accretion
spectrum is assumed to be flat, in erg cm$^{-2}$ s$^{-1}$ \AA$^{-1}$,
with wavelength for all filters longer than 4000 \AA, following
\citet{Herczeg2014}.  The bolometric correction to convert the excess
accretion continuum is calculated from the pure hydrogen slab models
developed by \citet{Valenti1993}. These slab models are less
physically realistic than shock models \citep{Calvet1998,Ingleby2013},
but both approaches provide similar bolometric corrections to obtain
the final accretion luminosity.   \par

These model spectra are then combined to reproduce the combined
optical/near-IR photometry using $\chi^2$ fits with two different
methods.  In the first set of fits, the photospheric luminosities are
fixed using JHK photometry and effective temperatures are fixed to the
values obtained from near-IR spectra (Table~\ref{tab:2}).
The only free parameter is the excess luminosity of the slab.  In the
second set of fits, the temperature, photospherivc luminosity, and
excess accretion luminosity are all treated as free parameters.  The
best fit effective temperatures and luminosities are listed in
Table~\ref{tab:2}. Radii and the photospheric luminosities of the
three object are calculated by scaling the BT-Settl synthetic spectra
to the observed SED. The uncertainty in radius is derived from the
uncertainties in extinction and distance. The range in acceptable mass
is set by fitting with pre-main sequence evolution model following
Fig.5 in \cite{Bowler2011}.

The spectral fits to DH~Tau~b and GSC 06214-0210 b are well matched to
most of the observed photometric points.   Possible K band excesses of
0.3-0.4 mag.~could be explained by emission from the warm disk,
similar to the L-band excess around GSC~06214-00210
\citep{Ireland2011}.  Alternatively, young substellar
and planetary mass objects tend to be systematically redder than model
spectra because of the gravity dependence of dust grain sizes
and the height of clouds \citep[e.g.][]{Marley2012}.

\begin{figure*}[!t]
  \centering
\plottwo{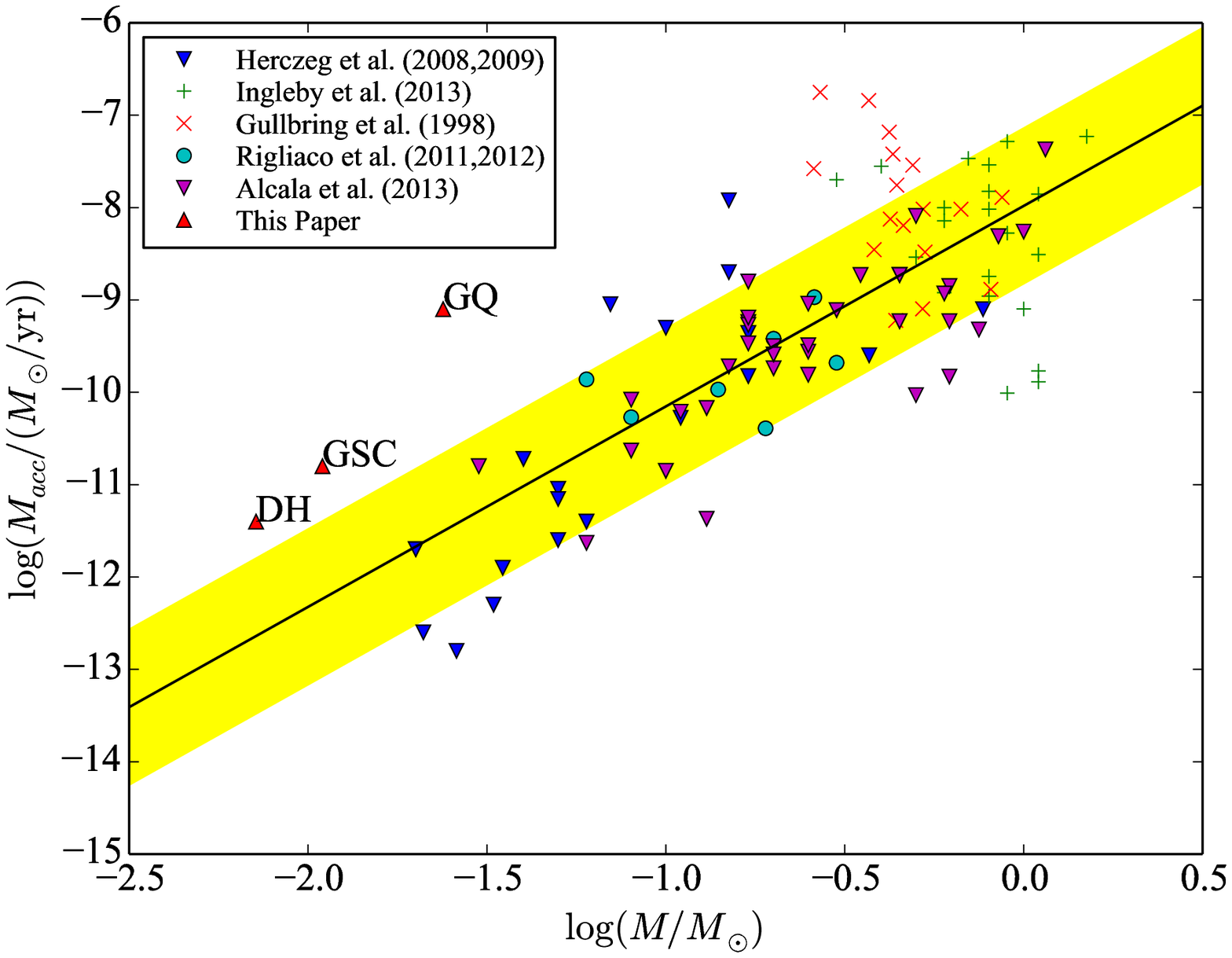}{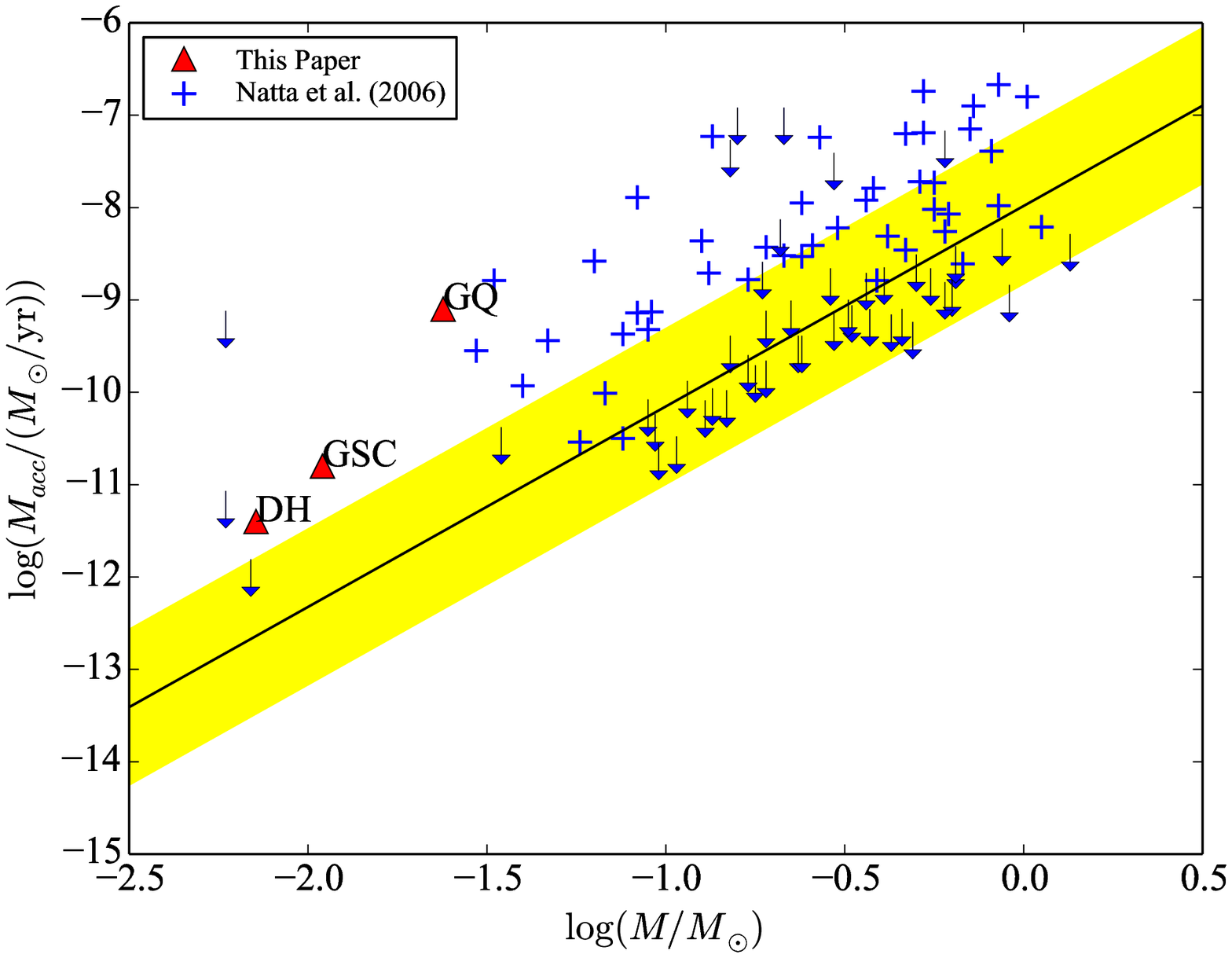}
\plottwo{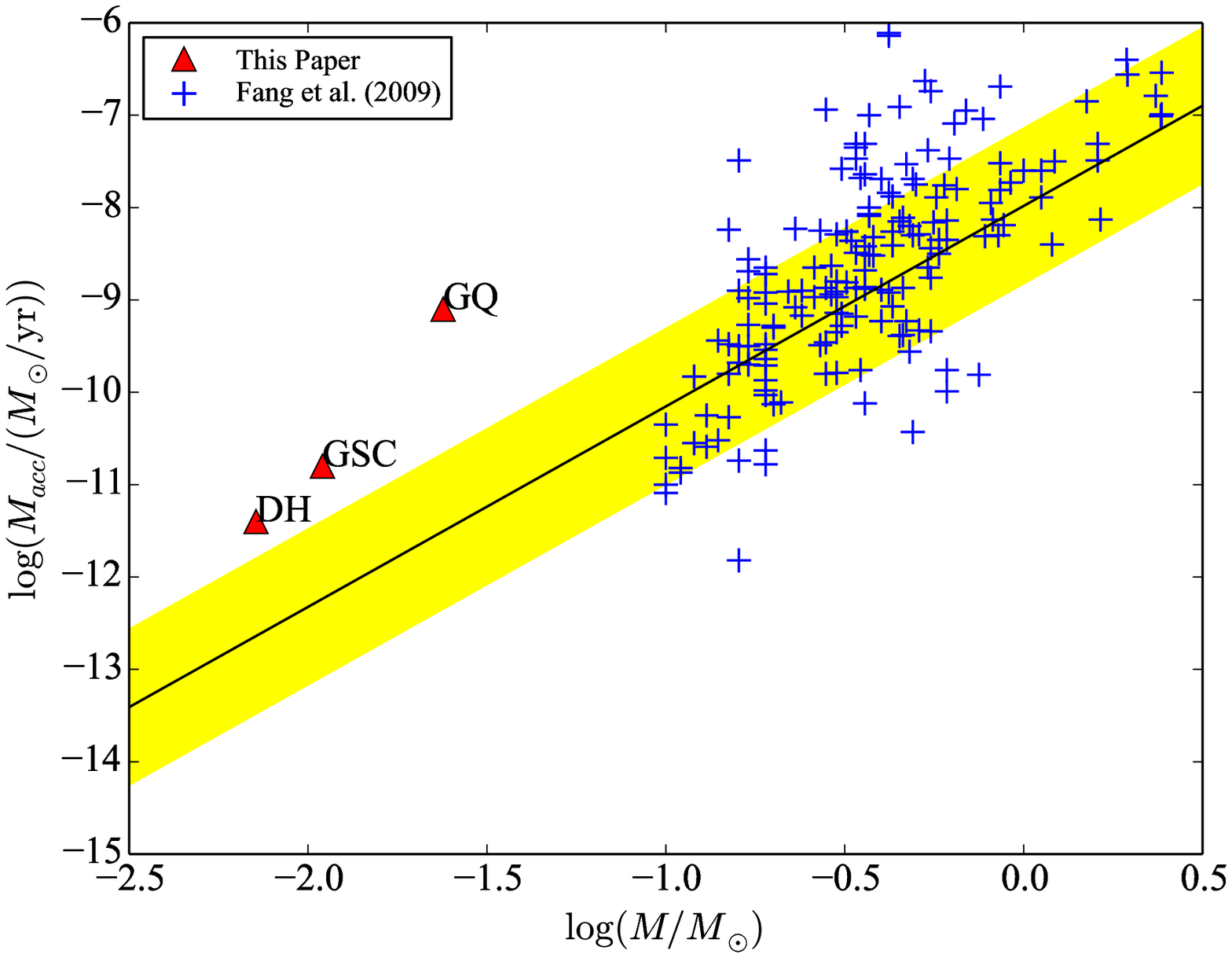}{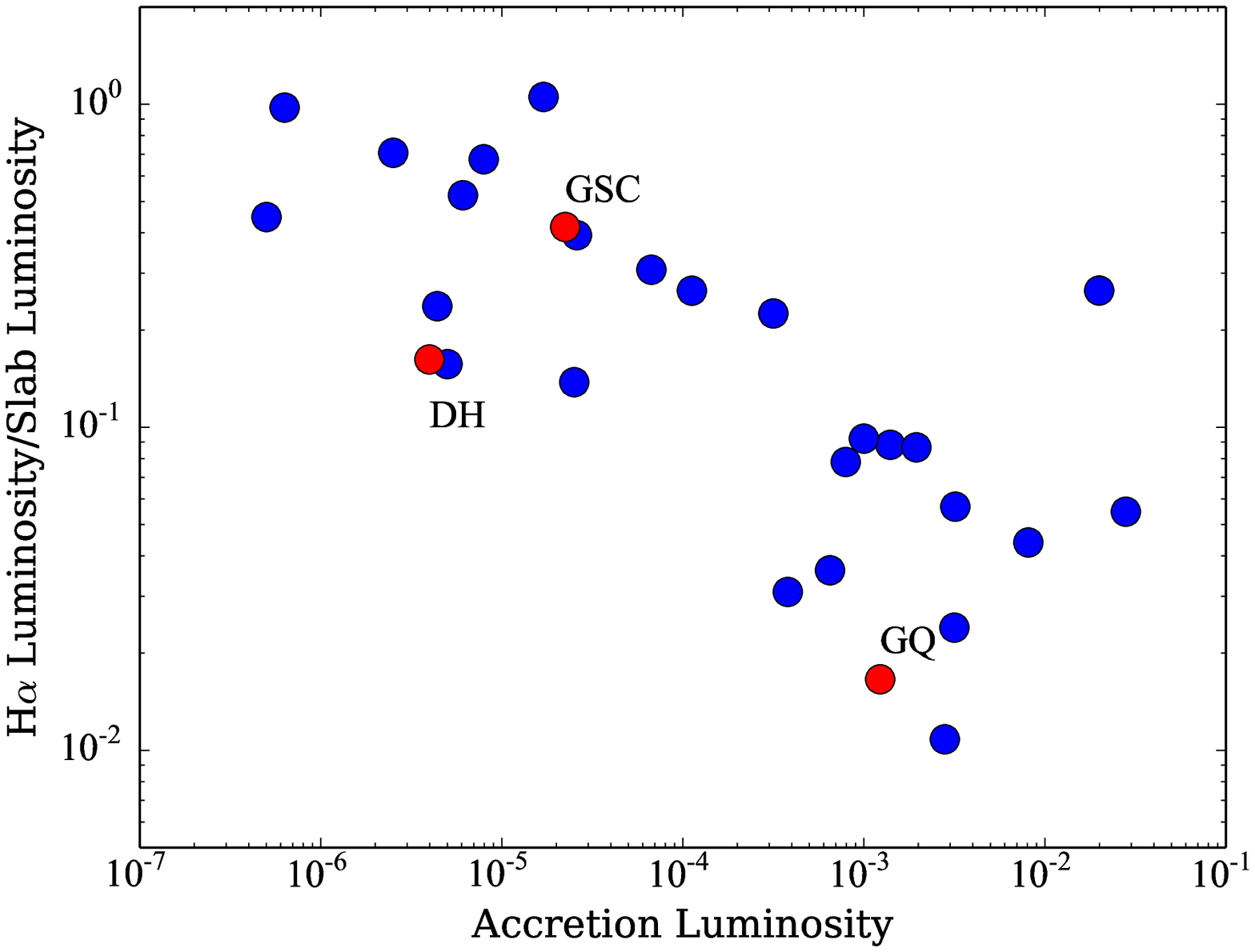}
\caption{Accretion properties of the planetary mass companions
  compared to other young accretors.  The three planetary mass companions have accretion rates that are one
  order of magnitude higher than expected from the correlation between
  object mass and UV excess accretion rates (upper left) and from
  accretion rates obtained from ombined measurements
  of the H$\alpha$, H$\beta$ and \ion{He}{1} $\lambda5876$ lines
  (lower left), but
  are consistent with accretion rates measured in 
$\rho$ Oph from Pa$\beta$ and Br$\gamma$ (upper right).  The yellow shaded region
shows the $\pm1\sigma$ region from best fit line between mass and
  UV excess accretion rates and demonstrates the offsets between
  different studies.  Lower right:  The ratio of H$\alpha$ luminosity to slab
  luminosity increases with lower accretion luminosity.}
\label{fig:3} 
\end{figure*}

The J-band and F850LP photometry of GQ Lup b have similar fluxes,
which is not expected from very cool objects and is poorly fit with the
model spectra. In fits with luminosity and temperature as free
parameters, no K-band excess is detected by the J-band flux is
severely overestimated by the models.
This discrepancy is unlikely to be explained by excess
line emission in the F850LP filter, because in that case more lines
would have been detected in the near-IR spectra. Large photometric
variability is often detected in accreting T Tauri stars
\citep[e.g.][]{Herbst1994} and may affect the
non-simultaneous comparison between the optical and near-IR
photometry.
The small separation between GQ Lup A and its companion
makes the primary star subtraction challenging and could introduce
photometric errors.

The effective temperatures measured here differ by a few hundred K
from the litearture values.  The fits of synthetic spectra to near-IR
spectra \citep[e.g.]{Patience2012} should yield more accurate
temperatures than our fits to optical+near-IR photometry.  However,
both temperatures are listed in case the depth of molecular bands is affected by
low gravity in ways that are not yet accounted for in the models, in
which case broadband SED fits would yeild better temperatures.
Spectral features may be weakened by any veiling from the disk in the
K-band and, for GQ Lup, possible J-band veiling from the accretion flow.
Accretion luminosities and rates are similar for both 
approaches and are listed separately in Table 1.

The total accretion luminosity, $L_{acc}$ is calculated by adding
excess continuum emission from the slab, $L_{slab}$ and the excess
line emission, in this case H$\alpha$, $L_{{\rm H}\alpha}$.  The
contribution of line emission to accretion rate measurements has been
ignored in all previous publications.  Following
\citet{Gullbring1998}, the accretion rate $\dot{M}$ is obtained by
\begin{equation}
  \dot{M}\approx\frac{1.25R_{*}L_{acc}}{GM_{*}} \label{eq:accL}
\end{equation}

The accretion luminosities and rates are listed in table \ref{tab:1}.
The extinction uncertainty, here assessed as $\sim 0.5$ mag.,
dominates the factor of $\sim 2$ uncertainty in accretion luminosity.
The factor of 3--5 uncertainty in accretion rate includes
uncertainties in the radius and temperature.  A detailed description
of errors is discussed in \citet{Herczeg2008}. The uncertainty in
extinction measurements introduces large errors in accretion
luminosity and rate. For GQ Lup b, an assumed extinction of
$A_{\mathrm{V}}=0.4$ mag. would yield $\log(L_{acc}/L_{\odot})=-3.7$ and
$\log(\mdot/(\mdotyr))=-10.1$.

\section{Discussion}

Accretion is detected in both the H$\alpha$ line and in excess optical
continuum emission from DH Tau b, GSC~06214-00210 b, and GQ Lup
b.  Accretion had been previously detected for all three objects in
Pa$\beta$ emission, although this emission was undetected in some
spectra of GQ Lup b.  For GSC~06214-00210 b, the accretion luminosity
measured here (when including H$\alpha$ luminosity) of $\log
\L_{acc}/L_\odot=-4.6\pm0.5$ is similar to the $-4.4\pm1.3$ measured
from Pa$\beta$ by \citet{Bowler2011}, although our direct measurements
have much smaller error bars.

The accretion rates calculated here demonstrate that these objects
have their own disks. The separations (100--300 AU) of the companion
objects to their primaries are too large to support such high
accretion rates. Moreover, there is no evidence to indicate the
presence of a primordial disk around the primary star GSC~06214-00210. 

These sources are at the brown dwarf-planet mass boundary, are among
the lowest mass sources with accretion measured directly from excess
optical spectra, and are the lowest mass companions to stellar primary
stars with measured accretion rates.  Previous detections of accretion
onto very low mass objects has been diagnosed through emission lines
\citep[e.g.][]{Bowler2011,Joergens2013}, which are indirect probes of
accretion rate, and directly from the Balmer jump for several
free-floating brown dwarfs \citep[e.g.][]{Herczeg2009,Rigliaco2012}.

Figure \ref{fig:3} shows correlations between mass and accretion
rate, as measured directly from excess Balmer continuum emission and
indirectly from line luminosities.  The accretion rates
measured here are more than one order of magnitude higher than
expected from the correlations obtained for accretion rates measured
directly from excess Balmer continuum emission \citep{Gullbring1998,Herczeg2008,Herczeg2009,Rigliaco2011,Rigliaco2012,Ingleby2013,Alcala2013} and separately for accretion rates
measured indirectly from optical lines \citep{Fang2009}.   This offset is not detected
when comparing our accretion rates to those measured for stars in $\rho$ Oph
from Pa$\beta$ and Br$\gamma$ \citep{Natta2006}.   However, the $\rho$ Oph
accretion rates at low object mass are anomalous relative to other studies, either because the objects are younger or because of
methodological differences.  Moreover, half of the stars in $\rho$ Oph
with disks have only upper limits in accretion luminosity.
While
differences in ages can complicate these comparisons, the high accretion rates are consistent with literature
values only if the objects have the same
age as $\rho$ Oph and are therefore younger than the stars in their
parent associations.  Although upper limits of our non-detections
should be considered, our program was complete for every known
planetary mass companion at the time of proposal submission and these
three objects are all high outliers in accretion rate.


These results suggest that wide planetary mass companions have higher accretion rates
than expected for objects formed from the fragmentation of a
protostellar core.  If correct, then models for their formation and evolution
models for the formation and evolution should account for these high
accretion rates and the survival of their disks.
Since core accretion is unlikely to form
planets at such large radii in the disk, the systematically high
accretion rates suggests either formation by gravitational instability
or different disk evolution between single planetary mass
objects/brown dwarfs versus those around stellar mass objects. 

In previous studies, the accretion rate has been calculated from only
the luminosity from the continuum emission. The H$\alpha$ luminosity
is usually negligible for solar mass stars. For GQ Lup b and DH Tau b,
the H$\alpha$ luminosity is ~5\% of the accretion continuum
luminosity. However, for GSC~06214-0210 b, H$\alpha$ luminosity is
equivalent to the total slab luminosity and accounts for ~50\% of the
total accretion luminosity.  Fig.~\ref{fig:3} shows the ratio of
H$\alpha$ luminosity to accretion continuum luminosity from our work
and from literature values
\citep{Herczeg2008,Herczeg2009,Rigliaco2011,Rigliaco2012}, when
measured simultaneously. The percentage of accretion luminosity that
escapes in H$\alpha$ increases with lower accretion luminosities,
possibly because of lower opacities and temperatures in the accretion
shock and accretion funnel flow.  Especially for low-mass objecst with
high $L_{{\rm H}_\alpha}/L_{slab}$ ratios, the accretion luminosities
may be underestimated if Ly$\alpha$ is significantly brighter than
H$\alpha$.  Unfortunately, Ly$\alpha$ emission from CTTSs is difficult to observe 
because of line-of-sight \ion{H}{1} absorption \citep{Schindhelm2012}.
Indeed, the one high point in $L_{{\rm H}_\alpha}/L_{acc}$ among solar
mass stars is TW Hya, which has an Ly$\alpha$ luminosity of $\sim
0.01$ $L_\sun$, or 0.5 $L_{slab}$ and 2 $L_{{\rm H}_\alpha}$
\citep{Herczeg2004}.

The strength of H$\alpha$ emission suggests an alternate method to
search for forming protoplanets.  The accretion continuum is spread
over a large wavelength region, so the contrast between the primary
and an accreting secondary star is still high.  However, if 10-50\% of
the accreting flux is produced in a single line, then the contrast
between the star and any accreting companion becomes much smaller.
Targeted searches for H$\alpha$ emission from companions may be a
powerful technique to find proto-Jupiters that are undergoing their
main phase of gas accretion. Indeed, \citet{Close2014} recently used
the new Magellan Adaptive Optics system to detect H$\alpha$ emission
and calculate accretion onto the faint companion around HD 142527.

\section*{acknowledgements}

We thank the anonymous referee for useful suggestions that helped to
clarify our analysis and results.  GJH is supported by the Youth
Qianren program in China. Based on observations associated with \HST{}
GO program 12507 and made with the NASA/ESA
Hubble Space Telescope, obtained at STScI, which is operated by AURA,
Inc., under NASA contract NAS 5-26555.

\end{document}